# AN APPROACH FOR AGILE SOA DEVELOPMENT USING AGILE PRINCIPALS

Majlesi Shahrbanoo, Mehrpour Ali and Mohsenzadeh Mehran

Department of Computer Engineering, Science and Research Branch, Islamic Azad University, Tehran, Iran

```
sh.majlesi@srbiau.ac.ir
a.mehrpour@srbiau.ac.ir
mohsenzadeh@srbiau.ac.ir
```

## ABSTRACT

*In dynamic and turbulent business environment, the need for success and survival of any organization is the ability of adapting to changes efficiently and cost-effectively. So, for developing software applications, one of the methods is Service Oriented Architecture (SOA) methodology and other is Agile Methodology. Since embracing changes is the indispensable concept of SOA development as well as Agile Development, using an appropriate SOA methodology able to adapt changes even during system development with the preservation of software quality is necessary. In this paper, a new approach consisted of five steps is presented to add agility to SOA methodologies. This approach, before any SOA-based development, helps architect(s) to determine Core Business Processes (CBPs) by using agile principals for establishing Core Architecture. The most important advantage of this approach according to the results of case study is possibility of embracing changes with the preservation of software quality in SOA developments.*

## KEYWORDS

*Agile SOA, Agile Development Methodology, Core Business Process (CBPs), Core Architecture*

## 1. INTRODUCTION

In today's competitive environment, because of customers' ever-changing requirements, using an appropriate software development methodology is one of the most critical issues. On one hand agile software development methodology encourages rapid and flexible response to changes by emphasis on customer involvement and its feedback, and also delivery of several small releases. On the other hand, the applying of SOA, as a pervasive strategy, for developing software application is increasing since it focuses on the ability to respond to changes [1]. Since adapting to changes is the indispensable concept of SOA development as well as Agile Development, it seems that using agile methodology is a natural fit to develop SOA applications, but they are fundamentally different and there is much debate about how they can be compatible, as SOA is a top-down approach while Agile is a bottom-up system development methodology. On the other hand, agile development methodologies don't act well against complexities which are the nature of SOA projects and cause applications with poor quality while one of the promises of SOA as an architectural style is to satisfy software quality. So, to profit from the advantages of both methodologies, more adaptable applications with higher quality, we need to embellish SOA development methodology with agile development principals. Thus this SOA methodology is able to adapt changes even during system development with the preservation of software quality.

In this paper, we propose an approach consisted of five steps in order to add agility to SOA methodologies. To achieve this, in this approach we have attempted to help architect(s) to establish a Core Architecture before any SOA-based development by applying of the most important principal of agile methodologies, customer involvement. To establish a Core Architecture, determining CBPs are necessary. CBPs are the business processes which influence on architectural decisions and thus on shaping the architecture, and Core Architecture is base for the whole of architecture in which software architect attempts to satisfy all of system quality attributes with making architectural decisions correctly. One of the most important advantages of establishing Core Architecture is embracing changes with the preservation of software quality.

Furthermore, delivering working software is another important principal of agile methodologies and proposed approach in order to support it, does requirements prioritization and selecting processes for current release appropriately with the nature of SOA projects. In section 2 in this paper, related work about SOA development are discussed. Section 3 describes proposed approach in detail. Section 4 shows the results gained by using proposed approach. Finally in the last section, the characteristics of proposed approach and future works have been concluded.

## 2. RELATED WORK

To support SOA-based software development, several SOA methodologies have been proposed. Many of these methodologies rely on business processes as primary inputs since business processes are suggested to be ideal candidates to represent the business requirements [2]. Service Oriented Modelling and Architecture (SOMA) [3] and Zimmerman's methodology [4] are the instances of these methodologies. In addition there are some methodologies such as Service Oriented Architecture Framework (SOAF) [5] which only provide a guideline without relying on special input. Another methodology is Service Oriented Unified Process (SOUP) [6] which tries uses of the best aspects of eXtreme Programming (XP) [7] [8], and Rational Unified Process [9] for an SOA project. This methodology is not documented completely, so it is not applicable.

Furthermore, there are several SOAD (Service Oriented Analysis and Design) approaches. The most famous of these approaches is the approach proposed by Thomas Erl [10]. The SOAD approaches proposed in [2] [11] are suitable for small and medium organizations. In spite of these methods and the efforts which have been done in order to a detailed method to develop SOA-based systems, SOA methodologies are not mature and there is not an approach which is broadly accepted. To ensure that SOA is defined and built using appropriate tools and methods is necessary [1]. Since embracing changes is the indispensable concept of SOA, it's seemed using agile methodology is natural fit to develop SOA-based systems. In this area, much debate has been done in [1] [12] [13][14][15][16][17] [18] [19], and despite consensus in the usefulness of an agile SOA methodology, there has not been considerable work in this area. Ervin et al. [20] despite expressing the immaturity of SOA methodologies, they have named the agility as one of the features of these methodologies and compared them from this point. Furthermore, the usefulness of XP for SOA projects has been examined in [16] [17] and in [21] integrating of SOA and Agile into a complementary partnership is considered. Also [22] explains how specific Agile practices support SOA development. The result shows that using agile development principals for developing SOA systems requires the adjustment of these principals with these kinds of projects.

To achieve greater agility in SOA development, agile methodologies are seemed to be fit to develop such systems. Although agile development methodologies are successful in dealing with changes, but they don't act well against complexities [23] which are the nature of SOA projects because of the lack of the pre-defined design of system. For developing each system, a structure or architecture is needed for better communication between stakeholders and when the system is larger and more complex, the architecture is required more. In other words, lack of architecture make the system unmanageable and by laps of time, the time and the cost of implementation and maintenance will increase. Also the quality which is required will not be met. So in the development of SOA which is an architectural style and uses services as building blocks to embrace changes in business environment by composing of services and creating composite services, a pre-thought of design is required. But on the other hand, Big Design Up Front (BDUF) is not also convenient and increases risk since the designing is being done when minimum knowledge of project requirements is obtained. Thus software architecture must be supposed and should be done in a manner which supports the development process agility. So a high level view of architecture and postponement of architectural decisions as much as possible at the time of implementation is a good compromise. For this purpose, Ambler who is the theoretician of agile modeling believes on the formation and modeling of architecture during iterations[24]. It means that at the start of development, we should be satisfied with a perspective of the architecture and then during the iterations, architecture should be completed concurrently with the evolution of requirements. So, with this conception the presence of agile SOA methodology is not only possible, but also more convenient because of services as the building blocks of architecture.

To realize such conception and make these two approaches compatible, SOA and agile methodologies, proposed approach helps system architect(s) to establish a Core Architecture before any SOA-based development with customer collaboration. To establish Core Architecture determining CBPs correctly is the most important issue since they are building blocks of Core Architecture. To determine CBPs, the quality attributes and priority of each business process are the foundation of this case. Quality attribute requirements are architecture driver, means that they shape the architecture [25] and they have an important role to determine CBPs. Fortunately the most important resource for determination of quality attribute requirements has been known, it is business goals[26], so system architect need them to develop Core Architecture. Clements et al. in [27][28][29] examined business goals in this point and the importance of them as architectural knowledge. Thus proposed approach which is consisted of five steps profits by business goals and customer involvement to determine quality attributes and priority of each business process. In the next section we describe how we can determine CBPs.

## 3. PROPOSED APPROACH

As mentioned before, the goal of proposed approach is to help architect(s) to establish Core Architecture before any SOA-based development. For this purpose, in this approach we determine quality attributes and priority of each business process by using business goal and in a manner in which different customers collaborate closely with architect(s). The architects need them in order to find out which business processes are CBP.

Before describing the proposed approach in detail, we need to define CBP exactly. CBP is a business process which is important to make architectural decisions and thus establish Core Architecture. It may be a high priority business process without any quality attribute or a high priority business process in which there are some quality attributes specially the ones with noticeable risk.

### 3.1. Prioritization of Business Goals

Before business goals prioritization, we need express business goals clearly. To achieve this, we can use business goals scenario which has been proposed in [25]. Because of the nature of SOA projects, business goals prioritization must be done in a manner in which all kind of stakeholders with different viewpoints and importance are supported. For this purpose, we assume that each stakeholder group has some representatives and receives an Importance Coefficient between 0 and 1 so that the total number of them is 1. Importance Coefficient shows the importance of each stakeholder group. If we suppose N as the number of business goals and K as the number of stakeholder groups, each representative allocates a number between 1 and N to each business goals according to its importance. The greater number shows more importance. So we can compute Importance Degree of each business goal as formula1.

$$\text{ImpDegBG}_i = \sum_{j=1}^{j=k} \text{the average of importance BG}_i \text{ in SG}_j * \text{ImpDeg SG}_j \quad (1)$$

In which BG shows business goal, $\text{ImpDegBG}_i$ shows the Importance Degree of $BG_i$ and $\text{ImpDegSG}_i$ shows the Importance Degree of i-th stake holder group. Then we compute the priority of each $BG_i$ which is between 0 and 1 as fomola2.

$$BG_i \text{ Priority} = \frac{\text{ImpDegBG}_i}{N} \quad (2)$$

This prioritization method is very simple and accurate since each business goal is related to one or more stakeholders and business goals are prioritized independently.

### 3.2. Mapping Business Goals to Business Processes

In this step, there are some business processes that allow some business goals to be satisfied. These business processes must be mapped to these business goals. Since it is a many-to-many map, we use a Support Coefficient which is between 0 and 1 in order to be distributed between business processes that allow a certain business goal to be satisfied. The sum of these Support Coefficients must be 1. Since we have the priority of business goals and their related business processes, we can compute the priority of each business process as fomula3:

$$BP_i \text{ Priority} = \sum_{j=1}^{j=N} SC_{ij} * BG_j \text{ Priority} \quad (3)$$

In which BP shows business process and $SC_{ij}$ shows the Support Coefficient of $BP_i$ in order to satisfy $BG_j$. In this step we prioritize business processes by using the priorities of business goals. So, it is possible to find the missing and also useless processes. It is one of the most important advantages of this kind of prioritization. Much accurate prioritization is another advantage since it is more facile and understandable for stakeholders to prioritize business goals instead of business processes.

### 3.3. Extraction of Quality Attribute Scenarios related to each Business Goals

In third step, for each business goals, quality attribute scenarios must be extracted by using a Goal Tree showed in figure1. Goal Tree is the same as Utility Tree in ATAM (Architecture Tradeoff Analysis Method)[30][31], with two differences. First, the root of Goal Tree is a business goal. Second, for each business goal we must have a separate Goal Tree. In Goal Tree, after root, there are some kinds of quality attribute in the second level such as performance, availability, security and etc. These quality attributes must be refine in next levels. Finally, the leaves of Goal Tree show the quality attribute scenarios of each business goal. The general scenario tables [30] can be used to suggest specific quality attribute scenario.

Using Goal Tree instead of Utility Tree is more helpful since the business goal in the root of Goal Tree will facilitate the extraction of quality attribute scenarios. Furthermore, it is quite clear since business processes which satisfy a certain business goal has been known in second step, the extraction of quality attribute scenarios will be done more facile for each business goals by involvement of stakeholders.

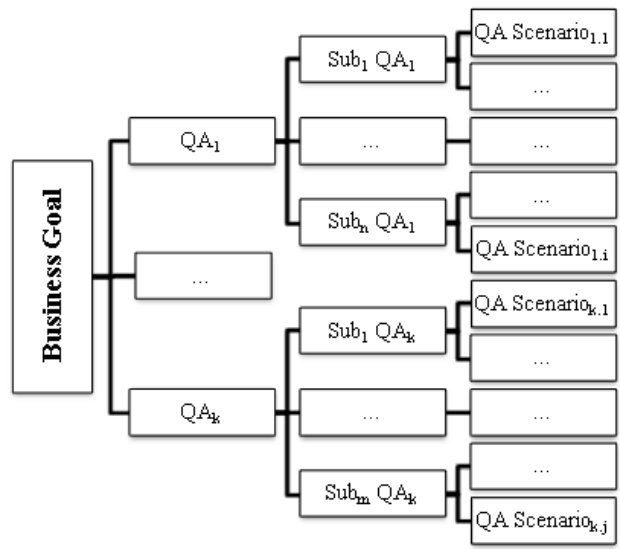

Figure1. Goal Tree to extract quality attribute scenario.

## 3.4. Determination of CBPs

This step is the most critical step of this approach in which architect(s) must determine CBPs according to the information obtained from previous steps. In this step, the priority of each business process and its quality attribute scenarios (from the relation between business process and its related business goals) are known. Also the risks of each quality attribute must be defined by architect(s). Now according to the information obtained and the definition of CBPs, there are some possible scenarios for each business process shown in table1. As can be seen, in two cases which are high light can certainly be said whether the business process is CBP or not. Business process is CBP if it has high priority and it has been mapped to some quality attributes with noticeable risk and it is not CBP if it has low priority and it has not been mapped to quality attributes. In other cases it depends to conditions in which architect(s) should decide. But most of time the results are the same as shown in table1.

Table1. Different scenarios related to each business process.

| Business Process | Mapped to Quality Attributes | | Not Mapped to Quality Attributes |
|---|---|---|---|
| | High Risk | Low Risk | |
| High Priority | Certainly CBP | Can be CBP | Can be CBP |
| Low Priority | Can't be CBP | Can't be CBP | Certainly Not CBP |

Besides this information, all of constraints like technical and business constraints [32] will influence on architectural decisions, so the knowledge and experience of architect(s) has a key role to establish Core Architecture.

## 3.5. Selecting Business Processes for Current Release

Now it's the time to develop Core Architecture and complete it during iterations. For achieve this, some business processes must be selected for current release. It is quite clear that some business processes which have the most priority should be selected but requirements are not independent from each other and they influence each other. So priority (business value) can't be the only criterion. Knowing the relations between requirements how they constraint each other is one of the most important factors [33]. Dependency factor is very important criterion which has not been supposed in much of prioritization technics [34] like the prioritization technique of XP in which only business value is supposed [35]. Thus proposed approach supposes this factor before selecting business processes for current release in order to fit SOA projects. To achieve this, business processes which depend on each other must be grouped so that the groups of business processes are independent. Then in each group the most dependent business processes must be combined as a new business process. Then business processes in each group must be placed in three categories:

1) Business processes which have high priority and high risk.

2) Business processes which have high priority and low risk.

3) Business processes which have low priority and low risk.

This categorization has been done as the same of agile development principal that selects the requirements with highest priority and risk for current release. So selecting business process for current release must be started from first category of each group and then continued with second and then third category. This method for selecting business processes for current release is compatible with the nature of SOA projects.

## 4. USING PROPOSED APPROACH

We applied proposed approach to a system being developed by Iran's global distribution which is a productions distributer organization. Shareholders, marketers and customers were supposed as three kinds of stakeholders who had different viewpoints and importances. The results gave us confidence that proposed approach is not only practical but also viable and valuable to develop agile architecture in an agile way since it is an easy approach to apply and also emphasizes on customer involvement. Furthermore knowing business processes and business goals at the start, five steps took only some hours.

## 5. CONCLUSIONS

In this paper, we have proposed an approach to add agility to SOA methodologies in order to profit from the advantages of both SOA and Agile developments. Proposed approach helps system architect(s) to establish a Core Architecture before the start of any SOA-based development by determining CBPs. For this purpose, our approach profits by the important role of business goals and also customer involvement. Also for delivering working software, this approach does business processes prioritization in a manner compatible with the nature of SOA projects since it supports different kinds of stakeholders with different viewpoints and importance and supposes business processes dependency. The results obtained from case study gave us confidence that this approach is practical to achieve these goals.

Although proposed approach is not dependent to a certain SOA methodology since it is being used before any SOA based development, but customizing it for a certain SOA methodology according to business processes and business goals as its inputs will have more advantages. So in the future we will focus on such approach.

## ACKNOWLEDGEMENTS

The authors would like to thank all those specially Dr. Ali Arsanjani, Paul Clements and J. D. Meier who have extended their support for successful completion of this work.